\documentstyle{article}
\begin{document}
\LARGE
\begin{center}

\vspace*{0.3in} \bf Inverse Temperature 4-vector in Special
Relativity

\vspace*{0.6in} \large \rm Zhong Chao Wu

Dept. of Physics

Zhejiang University of Technology

Hangzhou 310032, China

\vspace*{0.4in} \large  \bf Abstract
\end{center}
\vspace*{.1in} \vspace*{0.1in} \large

There exist several prescriptions for identifying the notion of
temperature in special relativity. We argue that the inverse
temperature 4-vector $\mbox{\boldmath $ \beta$}$ is the only viable
option from the laws of thermodynamics, and $\mbox{\boldmath $
\beta$}$ is a future-directed timelike 4-vector. Using a
superfluidity thought experiment, one can show that $\mbox{\boldmath
$ \beta$}$ is not necessarily along the time direction of the
comoving frame of the system, as is usually thought. It is
conjectured that, for an isolated system, the 4-vector is determined
from the entropy-maximum principle.

\vspace*{0.8in}\normalsize

PACS number(s): 05.70.-a, 03.30.+p, 05.90.+m, 67.10.-j


\vspace*{0.5in}

\pagebreak \large

No consensus has been reached in the treatment of thermodynamics in
the context of special relativity, even a whole century after the
formulation of special relativity [1]-[11].

In relativistic thermodynamics the most imminent problem concerns
the transformation laws of heat and temperature under the Lorentz
group. There are several published options in the literature:

(a)
\begin{equation}
\delta Q = \delta Q_0 \gamma^{-1},\;\;\;\; T  = T_0 \gamma^{-1},
\end{equation}

(b)
\begin{equation}
\delta Q = \delta Q_0 \gamma,\;\;\;\; T  = T_0 \gamma ,
\end{equation}

(c)
\begin{equation}
\delta Q = \delta Q_0,\;\;\;\; T  = T_0 ,
\end{equation}
where $\delta Q$ and $T$ represent heat and temperature
respectively, the variables with (without) subscript $0$ denote
those observed in the comoving (laboratory) frame, and $\gamma$ is
the Lorentz factor $ (1 - u^2)^{-1/2}$, where $\bf u$ is the
relative velocity of the comoving frame with respect to the
laboratory frame. In addition to options (a), (b) and (c), some
authors claimed (d) that ``there is no meaningful law of temperature
under boosts" [7][8].

Options (a), (b) and (c) are held by the authors of [1][2], [3][4]
and [5][6], respectively.  It is noted that, in principle, the
temperature in  (a) and (b) can be defined operationally using a
relativistic Carnot cycle [12]-[14].

One of the earliest attempts to find a covariant form of
thermodynamics was made by Israel and collaborators [9]-[11]. They
proposed a 4-vector $S^\mu$ for the flux of entropy, in a similar
way to the 4-vector for the flux of particle number. The particle
number in a comoving frame is a scalar. Likewise, we will show that
entropy in its comoving frame is a scalar as well, and so Israel's
proposal is supported. It is known that in a wide framework [15] a
path integral for a system in the Euclidean regime can be identified
by its partition function. The entropy of the system is the
logarithm of the partition function in a microcanonical ensemble.
For this ensemble the right representation should be chosen. In
particular, at the $WKB$ level, the entropy of the system is the
negative of its instanton action [16][17]. Since the path integral
and action are scalars, the entropy should be so too.

About other thermodynamic variables, various authors hold very
diversified opinions. In this letter we shall concentrate on the
temperature issue in special relativity. We shall use Planck units
in which $c = \hbar = k =G = 1$. The metric signature is $(- ,+, +,
+)$. For a system with a finite size the main obstacle in
formulating relativistic thermodynamics is the difference of true
and apparent transformations, and the calculation can be very
complicated [18]-[20]. This is due to the loss of an absolute
meaning of simultaneity in special relativity. To avoid the effect
of the finite size, we consider a continuous medium of infinite
size, or a medium of finite size but with a periodic boundary
condition.

Apparently, in the framework of special relativity, if one considers
the zeroth component of a 4-vector $\mbox{\boldmath $ \beta$ }$ as
 the inverse ``temperature" $T^{-1}$ and assumes that $ \beta_\mu \equiv u_\mu/T_0$,
where $u_\mu$ is the 4-velocity of the system, then $\mbox{\boldmath
$ \beta$ }$ has components $(T^{-1}_0, 0,0,0)$ in the comoving frame
[21][22], and we should obtain $T  = T_0 \gamma^{-1}$ in the
laboratory frame, in agreement with option (a) for the zeroth
component. If one takes the zeroth component of a 4-vector $\bf T$
as the ``temperature" $T$, and assume that $\bf T$ has component
$(T_0, 0,0,0)$ in the comoving frame, then we should easily obtain
$T  = T_0 \gamma$ in the laboratory frame, that agrees with the
opinion (b) for the zeroth component. Option (c) implies that the
temperature would be a scalar.

In the Israel covariant formulation of thermodynamics, not only the
equilibrium problem in the presence of gravity was investigated, but
also the off-equilibrium problem was studied, using the entropy flux
$S^\mu$ to reformulate the First Law. However, it was expected that
the controversy about the transformation law of thermodynamic
quantities would never lead anywhere [10].

Earlier van Kampen [21] considered the temperature as a scalar
$T_0$, and proposed a covariant form of the First Law, using
$\beta_\mu = u_\mu/T_0$. It seems that the inverse temperature
4-vector is redundant. Our proposal is distinct from his and other
similar arguments; in our case, one cannot always identify
$\beta_\mu$ as $u_\mu/T_0$, as mentioned earlier. All other
arguments on an inverse temperature 4-vector are based on the
existence of a rest-frame and that the vector $\mbox{\boldmath $
\beta$ }$ is $a\; priori$ oriented along the comoving 4-velocity of
the system. However, as was pointed out by Israel, the notion of a
rest-frame is not always well-defined, therefore Lorentz invariance
as applied to thermodynamics can not be devoid of physical content
[10].

Our proposal is distinct from these arguments. We argue that the
notion of temperature should be replaced by the inverse temperature
4-vector $\mbox{\boldmath $ \beta$ }$. It will be shown that the
inverse temperature 4-vector $\mbox{\boldmath $ \beta$ }$ is the
only viable option, with $\mbox{\boldmath $ \beta$ }$ being a
future-directed timelike 4-vector which, however, is not necessarily
along the time direction of the comoving frame. That is, there
always exists a frame in which $\mbox{\boldmath $ \beta$ }$ takes
the form $(T^{-1}_0, 0,0,0)$, but this frame is not necessarily
identical with the comoving one.

In a continuous medium the law of energy-momentum conservation reads
\begin{equation}
 T^{\mu \nu}_{\;\; ,\mu} = 0,
\end{equation}
where $T^{\mu \nu}$ is the total energy-momentum stress tensor. In
general, the conservation law includes the effects of both heat
exchange and applied work. The heat exchange can be considered as
the zeroth component of the heat vector. Its spatial components
represent the effect of the heatlike force [23].

Now let us consider a superfluidity thought experiment. Below some
critical temperature, liquid $^4He$, under thermal equilibrium
conditions, is capable of two different states at the same instant,
the normal and superfluid states [24][25]. The liquid $^4He$ model
and its generalized model have previously been studied by Israel
[10]. Here for simplicity, we only assume that there are two weakly
interacting components, and these two components mutually penetrate
without viscosity. Their energy-momentum is additive, that is, the
total energy-momentum of the medium is the sum of those of the two
components. This means that their interaction energy-momentum is
negligible, although the interaction between the two states still
exists. Each of the two components (states) has its own local
density $\rho_i$ and velocity $\bf v_{\rm i}$$(i = 1,2)$.

For our model (4) is rewritten as
\begin{equation}
\sum_{i} T^{\mu \nu}_{\;\;i, \mu} = 0.
\end{equation}
In addition to the energy-momentum tensor $T^{\mu \nu}_{\;\; i}$, in
general, there exist a number of 4-vectors $J^{\mu}_{Mi} $
representing the flux densities of conserved charges $M$. Their
conservation laws are expressed as
\begin{equation}
\sum_{i} J^{\mu}_{Mi , \mu} = 0.
\end{equation}
Following Israel, using the entropy 4-flux $S^{\mu}_i$, from the
First and Second Laws of thermodynamics one can write the following
covariant equation [10]
\begin{equation}
\sum_iS^{\mu}_{i , \mu } = - \sum_{i} \left ( \sum_{M}
\alpha_{Mi}J^{\mu}_{Mi ,\mu} +\beta_{\nu i} T^{\mu \nu}_{i
,\mu}\right ) \geq 0,
\end{equation}
where $S^{\mu}_{i , \mu }$ represents the creation rate of entropy
density, and
\begin{equation}
\alpha_{Mi} \equiv \mu_{Mi} |\mbox{\boldmath $ \beta$}_i|, \;\;\;\;
\end{equation}
where $\mu_{Mi}$ is the chemical potential of particle $B_{Mi}$,
which satisfies the equilibrium condition
\begin{equation}
\sum_M \alpha_M b_{Mi} = 0,
\end{equation}
where $b_{Mi}$ are the stoichiometric coefficients appearing in the
reaction equations
\begin{equation}
\sum_M b_{Mi} B_{Mi} = 0.
\end{equation}

It is important to emphasize that only the material part $T^{\mu
\nu}_{i(mat)}$ enters (7), ensuring that reversible flows of the
energy-momentum do not contribute to the entropy flux (this applies
to (12) and (15) below).

To avoid the effect of finite size, we use the differential form (7)
of the Israel formula for the creation rate of entropy density, in
which the term associated with the volume variation in the usual
formula vanishes. The trade off is, that for a given unit volume,
the particle number and other conserved charges are not fixed in the
process. Therefore, the macrocanonical ensemble and the fluxes of
the charges must be introduced.

If there is no interaction between the two motions, then each
component can itself be in thermal equilibrium and the entropy
creation rate vanishes [10]
\begin{equation}
S^\mu_{i , \mu} = 0,
\end{equation}

Now the interaction between the two components is switched on. In
general, the transportation of energy-momentum and other conserved
charges will increase the total entropy of the system. From (5)-(7)
one obtains
\begin{equation}
S^\mu_{\; , \mu}= \sum_{i} S^\mu_{i , \mu} = - \sum_{M}
(\alpha_{M1}- \alpha_{M2})\delta J^{\mu}_{M1 ,\mu} - (\beta_{\nu 1}
- \beta_{\nu 2}) \delta T^{\mu \nu}_{1 ,\mu} \geq 0,
\end{equation}
where $\delta J^{\mu}_{M1 ,\mu}$ and $\delta T^{\mu \nu}_{1 ,\mu}$
represent the arbitrary transfer of charges and energy-momentum from
component 2 to component 1. The Second Law demands each term in the
right hand side of (12) to be nonnegative. This means that the flux
$\delta J^{\mu}_{Mi ,\mu}$ is always transferred between components
from higher to lower chemical potential, as in the traditional
theory. In comparison with this, the heatlike flux $\delta
 T^{\mu \nu}_{1 ,\mu}$ behavior is more complicated than that in the
traditional scenario, in the latter heat is transferred from the
component with higher temperature to that with lower temperature.
Apparently, the necessary conditions for the two components to
approach equilibrium, i.e $S^\mu_{\; , \mu}=0$ are
\begin{equation}
\alpha_{M1}= \alpha_{M2}
\end{equation}
and
\begin{equation}
\beta_{\nu 1} = \beta_{\nu 2}.
\end{equation}

Eq. (13) was obtained by Israel [10]. Eq. (14) is the Zeroth Law of
thermodynamics in the new framework with the notion of the inverse
temperature 4-vector.

Since the comoving frames for the two components are different, the
same inverse temperature 4-vector cannot be along the two time
directions of both frames. One can conclude that, in general,
$\mbox{\boldmath $ \beta$}$ is not necessarily along the time
direction of the comoving frame of a system, as is usually believed.

If one accepts the notion of the inverse temperature 4-vector, then
the two states with non-vanishing relative velocity in the
superfluidity model can coexist in thermodynamic equilibrium, which
is distinct from thermal equilibrium in the special $^4He$ model
[10].

It seems that there exists an alternative approach. By using the
temperature 4-vector $\bf T$, the First Law can be recast into the
following covariant form
\begin{equation}
T^{\nu}S^\mu_{\;\; , \mu} = - \sum_{M} T^{\nu}\alpha_{M}J^{\mu}_{M
,\mu} + T^{\mu \nu}_{\;\; ,\mu},
\end{equation}
where $\alpha_M$ is redefined as
\begin{equation}
\alpha_M \equiv \frac{\mu_M}{|\bf T\rm |}.
\end{equation}

However, formula (15) is too restrictive. It is noted that here the
energy-momentum flux is along the orientation of the vector $\bf T$
instead  of its spacetime gradient if we temporally ignore the terms
of fluxes $J^\mu_{Mi}$. In contrast, in the traditional
thermodynamics, the heat flux is parallel to the temperature
gradient for isotropic media. Therefore, this prescription has to be
abandoned.

Some authors claimed that the temperature must be invariant with
respect to relative uniform motions [5][6]. Considering two
equilibrium identical bodies, which are in uniform relative motion,
they argued that the heat exchange can be carried out in the course
of smooth contact of the bodies and the flow would be  at right
angles to the motion. The observer attached to one body would judge
the temperature of the other body as lower, according to option (a).
From the usual relation between heat flow and temperature, heat
would be transferred to the other body. On the other hand, the
observation from the other body would be vice versa. This causes
contradiction. The situation is similar for option (b). Therefore,
one has to adopt option (c).

The reason leading to the above consequence is that the First Law
was not treated in a covariant way. Roughly speaking, since the
entropy is a scalar and the heat flux is a vector [23], the
temperature must obey a 4-vector form.

It is concluded that the relativistic formulation of the First Law
demands the notion of the inverse temperature 4-vector
$\mbox{\boldmath $ \beta$ }$, which should take the role of the
traditional scalar temperature in classical thermodynamics. How to
measure its spatial components is another problem, since the
relative speed in the laboratory is much smaller than the speed of
light. Its effects might be found in relativistic astrophysics [26].

Let us turn to relativistic statistics. It is known that in the
comoving frame the Maxwell probability distribution for one-particle
velocity of an ideal gas is expressed as
\begin{equation}
f_M(\bf v\rm ;m, |\mbox{\boldmath $ \beta$}|) =[m|\mbox{\boldmath $
\beta$}|/(2\pi)]^{3/2} \exp (-|\mbox{\boldmath $ \beta$}| m\bf
v^{\rm 2}\rm /2),
\end{equation}
where $|\mbox{\boldmath $ \beta$}| = T^{-1}_0$, $m$ is the mass of
the particle, $\bf v \rm$ is its 3-velocity.

Its relativistic version was proposed by Juettner as follows [27]
\begin{equation}
f_J(\bf v\rm ; m, |\mbox{\boldmath $ \beta$}|)= m^3\gamma(\bf v
\rm)^5 \exp [-|\mbox{\boldmath $ \beta$}| m \gamma (\bf v \rm)]/Z_J,
\end{equation}
where $Z_J = Z_J(m, |\mbox{\boldmath $ \beta$}| )$ is the
normalization constant. In the laboratory frame, the Juettner
function becomes
\begin{equation}
f_J(\bf v^\prime \rm ; m, |\mbox{\boldmath $ \beta$}|, \bf u \rm)=
m^3\gamma(\bf v^\prime \rm)^5 \gamma (\bf u \rm)^{-1} \exp
[-|\mbox{\boldmath $ \beta$}| m \gamma (\bf u \rm)\gamma (\bf
v^\prime \rm)(1+ \bf u \cdot v^\prime)\rm ]/\rm Z_J,
\end{equation}
where $\bf u$ is the relative velocity of the laboratory with
respect to the comoving frame and  $\bf v^\prime$ is the particle
velocity in the laboratory frame. The extra factor $\gamma (\bf u
\rm)^{-1}$ is due to Lorentz contraction in  velocity space.
$f_J(\bf v^\prime \rm ; m, |\mbox{\boldmath $ \beta$}| , \bf u \rm)$
can be rewritten as
\begin{equation}
f_J(\bf v^\prime \rm ; m,  \mbox{\boldmath $ \beta$} \rm)=
m^3\gamma(\bf v^\prime \rm)^5 \gamma (\bf u \rm)^{-1} \exp [
\beta^\mu \epsilon_\mu(\bf v^{\rm \prime}) ]/\rm Z_J,
\end{equation}
where $\mbox{\boldmath $ \epsilon$}\rm(\bf v^{\rm \prime})$ is the
energy-momentum of the particle. If one accepts the notion of
inverse temperature 4-vector $\mbox{\boldmath $ \beta$}$, then the
exponent of the probability density function has covariant form. In
general, the Boltzmann factor in a Gibbs state should take the same
covariant form [22].

From (20) it follows that $\mbox{\boldmath $ \beta$}$ must be a
timelike future-directed 4-vector, otherwise the distribution (20)
cannot be normalized.

The Juettner distribution function (18)-(20) revised for
2-dimensional spacetime has been confirmed by numerical simulations
very recently [28].

One might ask what orientation  it should take. Our conjecture is as
follows: The entropy of an isolated system is a function of
temperature and other thermodynamic parameters. Under the same
restrictions, the direction of the 4-vector is oriented in a way so
that the entropy takes a maximum value.

In this letter we dealt with the modest problem: the notion of
temperature in special relativity. The notion of temperature in
general relativity is much more complicated [4], since one has to
consider the group of general coordinate transformations, instead of
the Lorentz group. Firstly, in the classical framework $(\hbar = 0)$
there does not exist a local definition of gravitational
energy-momentum. Secondly, in the quantum framework, there exist
fluctuations in quantum fields [29]. In particular, there does not
exist an unique vacuum state even in the non-inertial frame of
Minkowski spacetime [30], let alone in a curved spacetime.

 \large \vspace*{0.1in} \rm

\bf Acknowledgement: \vspace*{0.1in} \rm

This work is supported by NSFC No.10703005 and No.10775119.

\large \vspace*{0.1in} \bf References:

 \rm

[1] EINSTEIN A., \it Ueber das Relativitaetsprinzip und die aus
demselben gezogenen Folgerungen. Jahrb. Rad. u. Elektr., \rm \bf 4
\rm (1907)411.

[2] PLANCK M., \it Ann. d. Phys., \rm \bf 26 \rm (1908) 1.

[3] OTT H., \it Zeitschr. d. Phys., \rm \bf 175 \rm (1963) 70.

[4] ARZELIES H., \it Nuov. Cim., \rm \bf 35 \rm (1965) 792.

[5] LANDSBERG P.T., \it Nature, \rm \bf 212 \rm (1966) 571.

[6] LANDSBERG P.T., \it Nature, \rm \bf 214 \rm (1967) 903.

[7] LANDSBERG P.T. and MATSAS G.E.A., \it Phys. Lett., \rm \bf A
223\rm (1996) 401.

[8] SEWELL G. L., \it J. Phys., \rm \bf A 41 \rm, (2008)382003.

[9] ISRAEL W., \it Physica, \rm \bf 204 \rm (1981) 204.

[10] ISRAEL W., \it J. Non-Equilib. Thermodyn., \rm \bf 11 \rm
(1986) 295.

[11] ISRAEL W. and STEWART J.M., in \it HELD, A. (ed.) Genaral
relativity and gravitation Vol. \bf 2\rm, \rm edited by HELD A.
(Plenum Press, New York) (1980).

[12] VAN LAUE M., \it Die Relativitaetstheorie, Vol. \bf I \rm
Vieweg, Braunschweig (1951).

[13] TOLMAN R.C., \it Relativity, Thermodynamics and Cosmology, \rm
Dover, N.Y. (1987).

[14] REQUARDT M., gr-qc/0801.2639v2 preprint, 2008.

[15] GIBBONS G.W. and HAWKING S.W., \it Phys. Rev., \rm \bf D \rm
\bf 15 \rm (1977) 2752.

[16] WU Z.C., \it Int. J. Mod. Phys., \bf D 6 \rm (1997) 199.

[17] WU Z.C., \it Phys. Lett., \bf B 659 \rm (2008) 891.
gr-qc/0709.3314.

[18] FERMI E., \it Nuov. Cim., \rm \bf 25 \rm (1923) 159.

[19] ROHRLICH F., \it Nuov. Cim., \bf XLV B \rm (1966) 76.

[20] BALESCU R., \it Physica, \rm \bf 40 \rm (1968) 309.

[21] VAN KAMPEN N. G., \it Phys. Rev., \bf 173 \rm (1968) 295.

[22] OJIMA I., \it Lett. Math. Phys., \rm \bf 11 \rm (1986) 73.

[23] RINDLER W., \it Introduction to Special Relativity,\rm
(Clarendon Press, Oxford) 1991.

[24] LONDON F., \it Nature, \rm \bf 141 \rm (1938) 643.

[25] TISZA L., \it Nature, \rm \bf 141 \rm (1938) 913.

[26] CARTER B. and SALUELSSON L., \it Class. Quant. Grav., \rm \bf
23 \rm (2006) 5367.

[27] JUETTNER F., \it Ann. Phys. \rm (Leipzig), \bf 34 \rm (1911)
856.

[28] CUBERO D., CASADO-PASCUAL J., DUNKEL J., TALKNER P. and HAENGGI
P., \it Phys. Rev. Lett., \rm \bf 99 \rm (2007) 170601.

[29] HAWKING S.W., \it Commun. Math. Phys., \rm\bf 43 \rm (1975)
199.

[30] UNRUH W., \it Phys. Rev., \rm\bf  D 14 \rm (1976) 870.

\end{document}